\documentstyle[prd,aps,floats,psfig]{revtex}
\draft
\begin{document}

\wideabs{
\vspace{-1.2cm}
\title{
A High Statistics Search for $\nu_e(\overline\nu_e) \rightarrow
\nu_\tau(\overline\nu_\tau)$ Oscillations}
\author{
        D.~Naples$^{4,\dagger}$, A.~Romosan$^{2,\dagger\dagger}$, 
        C.~G.~Arroyo,$^2$ L.~de~Barbaro,$^5$
        P.~de~Barbaro,$^7$ A.~O.~Bazarko,$^2$ R.~H.~Bernstein,$^3$
        A.~Bodek,$^7$ T.~Bolton,$^4$ H.~Budd,$^7$ J.~Conrad,$^2$
        R.~B.~Drucker,$^6$ D.~A.~Harris,$^7$ R.~A.~Johnson,$^1$
        J.~H.~Kim,$^2$ B.~J.~King,$^2$ T.~Kinnel,$^8$ M.~J.~Lamm,$^3$
        W.~C.~Lefmann,$^2$ W.~Marsh,$^3$ K.~S.~McFarland,$^7$
        C.~McNulty,$^2$ S.~R.~Mishra,$^2$ 
        P.~Z.~Quintas,$^2$ W.~K.~Sakumoto,$^7$ H. Schellman,$^5$
        F.~J.~Sciulli,$^2$ W.~G.~Seligman,$^2$ M.~H.~Shaevitz,$^2$
        W.~H.~Smith,$^8$ P.~Spentzouris,$^2$ E.~G.~Stern,$^2 $
        M.~Vakili,$^1$ U.~K.~Yang,$^7$ and J.~Yu$^3$
}
\address{
$^1$ University of Cincinnati, Cincinnati, OH 45221 \\
$^2$ Columbia University, New York, NY 10027 \\
$^3$ Fermi National Accelerator Laboratory, Batavia, IL 60510 \\
$^4$ Kansas State University, Manhattan, KS 66506 \\
$^5$ Northwestern University, Evanston, IL 60208 \\
$^6$ University of Oregon, Eugene, OR 97403 \\
$^7$ University of Rochester, Rochester, NY 14627 \\
$^8$ University of Wisconsin, Madison, WI 53706 \\
$\dagger$ present address: University of Pittsburgh, Pittsburgh, PA 15260 \\
$\dagger\dagger$ present address: Lawrence Berkeley National Lab, Berkeley, CA 94720 \\
}

\date{\today}
\maketitle
\begin{abstract}

We present new limits on $\nu_e (\overline{\nu}_e) 
\to \nu_\tau (\overline{\nu}_\tau)$ 
and $\nu_e(\overline{\nu}_e) \to \nu_s$ oscillations by searching
for $\nu_e$ disappearance in the high-energy wide-band CCFR
neutrino beam. Sensitivity to $\nu_\tau$ appearance comes 
from $\tau$ decay modes in which a large fraction 
of the energy deposited is electromagnetic.
The beam is composed primarily of 
$\nu_\mu$($\overline{\nu}_\mu$)
but this analysis uses the $2.3$\% $\nu_e$($\overline{\nu}_e$) 
component of the beam.
Electron neutrino energies range from 30 to 600 GeV and 
flight lengths vary from 0.9 km to 1.4 km. This limit improves
the sensitivity of existing limits for $\nu_e \to \nu_\tau$
at high $\Delta m^2$ and 
obtains a lowest 90\% confidence upper limit in $\sin^2 2\alpha$
of $9.9 \times 10^{-2}$ at $\Delta m^2 \sim 125$~${\rm eV^2}$. 
\end{abstract}

\pacs{PACS numbers: 14.60.Pq, 13.15.+g}
\twocolumn
}

Neutrino oscillations occur if neutrinos have non-zero mass and mixing. 
If recent evidence \cite{sk,lsnd} 
for neutrino oscillations is confirmed, it will radically 
alter our understanding of both particle physics
and cosmology. Neutrino oscillations may also explain the 
observed deficit of neutrinos from the sun. In the two-generation 
mixing formalism, the oscillation probability is given by
\begin{equation}
P(\nu_1 \rightarrow \nu_2) = \sin^2 2\alpha \sin^2 \left(\frac{1.27
\Delta m^2 L}{E_\nu}\right)
\label{eq:posc}
\end{equation}
where $\Delta m^2$ is the mass squared difference of the mass
eigenstates in ${\rm eV^2}$, $\alpha$ is the mixing angle, $E_\nu$ is
the incoming neutrino energy in GeV, and $L$ is the distance between
the point of creation and detection in km.

While the high $\Delta m^2$ regions of parameter space for
$\nu_\mu \to \nu_e$ and $\nu_\mu \to \nu_\tau$ oscillations
have been excluded to very low
mixing angles ($2\times10^{-3}$ for  $\nu_\mu \to \nu_e$
and $5\times10^{-3}$ for $\nu_\mu \to \nu_\tau$) the high $\Delta m^2$
$\nu_e \to \nu_\tau$ parameter space is much less constrained
as a result of the difficulty in producing high energy $\nu_e$ beams. 
The CCFR sample of over 20,000 charged-current $\nu_e$ interactions
comprises the largest sample of high energy $\nu_e$'s to date.   
Previous high-statistics limits were obtained 
from reactor experiments \cite{gos,bug} with much lower beam energies. 
Accelerator limits were obtained by BEBC \cite{bebc} 
and Fermilab E531 \cite{E531} which searched for $\nu_\tau$ appearance
in emulsion.

We previously reported a limit on $\nu_\mu \rightarrow \nu_e$
oscillations by searching for $\nu_e$ appearance \cite{alex} in the 
$\nu_e N$ charged-current data sample. 
In this report we present new limits using the same data sample
on $\nu_e \rightarrow \nu_\tau$ and $\nu_e \rightarrow \nu_s$ 
oscillations. Both limits use a $\nu_e$ disappearance test.
The $\nu_e \rightarrow \nu_\tau$ limit is also sensitive to 
$\nu_\tau$ appearance through $\tau$ decay modes 
in which a large fraction of the energy deposited 
is electromagnetic.

The CCFR detector \cite{ws90,bk91} consists of an 18~m long, 690~ton
target calorimeter with a mean density of ${\rm 4.2~
g/cm^3}$, followed by an iron toroidal spectrometer. The
target consists of 168 steel plates, each ${\rm 3 m \times 3 m \times
5.15 cm}$, instrumented with liquid scintillator counters placed
every two steel plates and drift chambers spaced every four plates.
The separation between scintillation counters corresponds to 6
radiation lengths, and the ratio of electromagnetic to hadronic
response of the calorimeter is $1.05$. The toroid spectrometer is not
directly used in this analysis which is based on the shower profiles
in the target-calorimeter.

\begin{figure}
\centerline{\psfig{figure=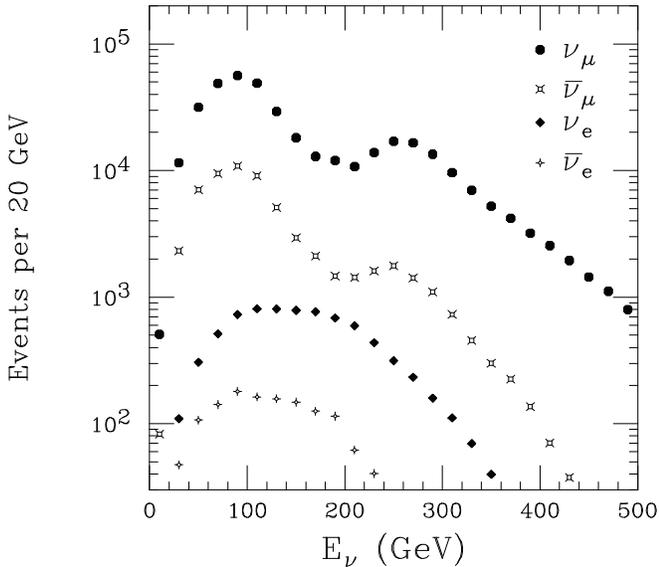,width=\columnwidth}}
\caption{Neutrino energy spectra for $\nu_\mu$, $\overline{\nu}_\mu$,
$\nu_e$, and $\overline{\nu}_e$ at the CCFR detector for the Fermilab
wide-band neutrino beam.}
\label{fig:beam}
\end{figure}

The Fermilab Tevatron Quadrupole Triplet neutrino beam
is created by decays of pions and kaons produced when $800$~GeV
protons hit a production target 1.4~km upstream of the neutrino detector.
The resulting neutrino energy spectra for $\nu_\mu$, $\overline{\nu}_\mu$,
$\nu_e$, and $\overline{\nu}_e$ are shown in Figure \ref{fig:beam}.
The $2.3$\% $\nu_e$ component of the beam used in this analysis 
is produced mainly from $K^\pm \rightarrow \pi^0 e^\pm
\stackrel{_{(-)}}{\nu_e}$ occuring in the 0.5~km decay region just downstream
of the production target. 
The $\nu_\tau$ content of the beam is less than $10^{-5}$.

Neutrino interactions observed in the detector can be divided into
three classes depending on the type of incoming neutrino and
interaction:
\begin{enumerate}
\item  $\nu_{\mu}N \rightarrow \mu^-X$ ($\nu_{\mu}$ charged-current (CC)
events).
\item $\nu_{e,\mu}N \rightarrow \nu_{e,\mu}X$ ($\nu_{e,\mu}$ neutral-
current (NC) events).
\item $\nu_e N \rightarrow eX$ ($\nu_e$ CC events).
\end{enumerate}

The majority (97.7\%) of events observed in the detector are
produced by muon neutrino interactions. 
The $\nu_\mu$~CC events can be identified by the presence of a muon
in the final state which penetrates beyond the end of the
hadron shower, depositing energy characteristic of a minimum ionizing
particle \cite{ws90} in a large number of consecutive scintillation
counters. Conversely, the electron produced in a $\nu_e$~CC event
deposits energy in a few counters immediately downstream of the
interaction vertex and is typically much shorter than the hadron 
shower. The separation of $\nu_{e,\mu}$~NC from the $\nu_e$~CC events
is accomplished by using the difference in 
energy deposition pattern within the shower region; the $\nu_e$~CC events
have a larger fraction of their energy 
deposited near the shower vertex.

In this analysis,
the three most important experimental quantities calculated for each
event are length, visible energy, and shower energy deposition 
profile. Event length is
determined to be the number of scintillation counters spanned from the
event vertex to the last counter with greater than
a minimum-ionizing pulse
height. The visible energy in the calorimeter, $E_{vis}$, is obtained
by summing the energy deposited in scintillation counters from the
interaction vertex to five counters beyond the end of the shower. The
shower energy deposition profile is characterized by the ratio of the
sum of the energy deposited in the first three scintillation counters
to the total visible energy. Accordingly, we define
\begin{equation}
\eta_3 = 1 - \frac{E_1 + E_2 + E_3}{E_{vis}} \label{eq:eta3}
\end{equation}
where $E_i$ is the energy deposited in the $i^{th}$ scintillation
counter downstream of the interaction vertex.

The event length is determined by the end of the hadron
shower for $\nu_\mu$~NC and $\nu_e$~CC events but is determined by the
muon track for most $\nu_\mu$~CC events. 
To isolate events without a muon track
we parameterize the event length as a function of energy for
which 99\% of hadron showers are contained as
\begin{equation}
L_{NC} = 4.0 + 3.81\times \log(E_{vis}).
\end{equation}
Events which deposit energy over an interval less than $L_{NC}$ 
counters are classified as ``short'', otherwise they are ``long''. 
The long event sample consists almost exclusively of class~1 events, while 
the short sample is a mixture of class~2, class~3, and class~1 events with a
low energy muon.

Events were selected with at least 30 GeV deposited in the
target calorimeter to ensure complete trigger efficiency.
Additionally, we require the event vertex to be
at least five counters from the upstream end and 
more than $L_{NC}+5$ counters from the downstream end of the target,
and less than $127$ cm from the detector center-line. The resulting data sample
consists of 632338 long events and 291354 short events.

To directly compare the long and short events a
muon track from the data was added to the short events to compensate
for the absence of a muon in NC events. The fraction, {\em f}, of
$\nu_\mu$~CC events with a low energy muon contained in the short
sample (which will now contain two muon tracks) was determined from 
a Monte Carlo simulation to be approximately 20\%.
A simulated sample of such events was obtained by choosing a long event
with the appropriate energy distribution from the data and combining it
with a second short muon track. The length of the short
track and its angular distribution were obtained from a Monte Carlo
sample of $\nu_\mu$~CC events.

A sample of $\nu_e$~CC interactions with a muon track added were obtained
by convolving an electromagnetic shower generated using GEANT \cite{g321} 
with an event from the long data sample with the appropriate
energy. This assumes $\nu_\mu - \nu_e$ universality. 
The energy distribution of $\nu_e$'s and the fractional
energy transfer $y$ were obtained from Monte Carlo. 
Because the hadron showers in the long data
sample already have a muon track, the $\nu_e$~CC sample can be compared
directly with the short and long events.

\begin{figure}
\centerline{
\psfig{figure=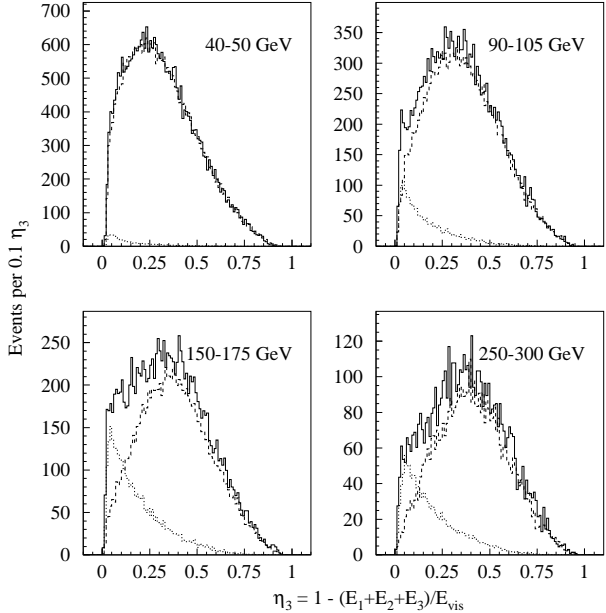,width=\columnwidth}}
\caption{$\eta_3$ distributions for short (solid line), long (dashed line),
and $\nu_e$~CC (dotted line) events in four of the energy bins studied.
The $\nu_e$~CC and long distributions are normalized to the respective
number of events predicted by the fit.}
\label{fig:etas}
\end{figure}

The long and short $\eta_3$ distributions were further corrected by
subtracting contamination due to cosmic ray events. The cosmic ray
background was estimated from an event sample collected during a 
beam-off gate using an analysis procedure identical to the one used for 
the data gates.
Additionally, the $\eta_3$ distribution of short $\nu_\mu$~CC events,
normalized to the predicted fraction {\em f}, was subtracted from the
short event sample. The $\eta_3$ distributions for short, long, and
$\nu_e$~CC events for various energy bins are shown in
Figure~\ref{fig:etas}.

We extract the number of $\nu_e$~CC events in each of 15 $E_{vis}$
bins by fitting the corrected shape of the observed $\eta_3$ distribution
for the short sample to a linear combination of long and $\nu_e$~CC
$\eta_3$ distributions:
\begin{equation}
\eta_3(short) = \alpha~\eta_3(long) + \beta~\eta_3(\nu_e CC)
\end{equation}
The $\chi^2$ of the fit in each of the 15 $E_{vis}$ bins ranges from
33 to 78 for 41 degrees of freedom (DoF) with a mean value of 48. 

\begin{figure}
\centerline{
\psfig{figure=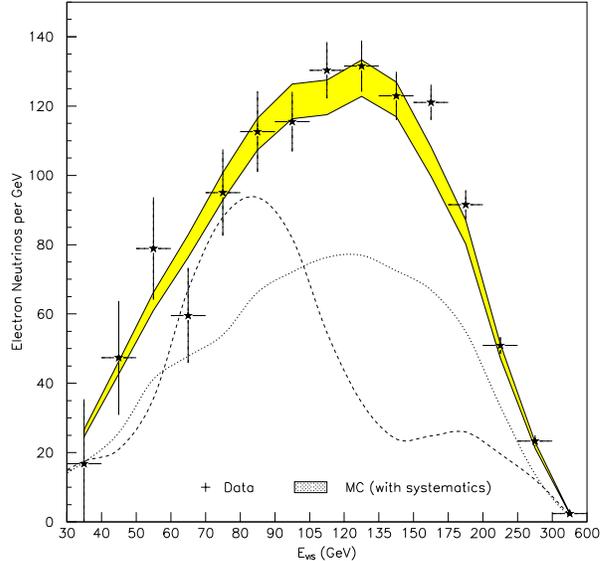,width=\columnwidth}}
\caption{Number of electron neutrinos as a function of visible energy.
For electron neutrinos the visible energy is equal to the total
neutrino energy. The filled band shows Monte Carlo prediction assuming
no oscillations. The curves shown are the effect of 
$\nu_e \rightarrow \nu_\tau$ oscillations for $\sin^2 2 \alpha = 1$
and $\Delta m^2 = 150$~${\rm eV^2}$ (dashed) and
$\Delta m^2 = 10000$~${\rm eV^2}$ (dotted).}
\label{fig:result}
\end{figure}

To search for $\nu_e$ oscillations the measured absolute flux of $\nu_e$'s
at the detector was compared to the flux predicted by a detailed
beamline simulation \cite{ca94}.
Figure \ref{fig:result} shows the measured number of
$\nu_e$~CC's for each energy bin compared with the predicted flux.
The $\chi^2$ value with a no-oscillations assumption is $6.8/15$ DoF.
We interpret a deficit in the measured $\nu_e$ flux as 
$\nu_e \rightarrow \nu_\tau$ (or $\nu_e \rightarrow \nu_s$) oscillations. 
($\nu_e \rightarrow \nu_\mu$ oscillations are excluded above mixing angles 
of $2\times10^{-3}$ at the 90\% confidence level in this $\Delta m^2$ range).

If $\nu_e \rightarrow \nu_\tau$ oscillation occurs,
some fraction, $f_{e}$, of $\nu_\tau$ charged-current interactions
will be observed in our $\nu_e$ data sample. These $\nu_e$~CC-like 
events result from $\nu_\tau$ interactions in
which a large fraction of energy deposited by the final 
state $\tau$ is electromagnetic.
To determine this fraction we simulated 
charged-current $\nu_\tau$ interactions in our detector using 
GEANT and a combination of LUND \cite{lund} to generate charged-current 
neutrino interactions and TAUOLA \cite{tauola}
to simulate tau lepton decays.
We fit the resulting $\nu_\tau$ charged-current Monte Carlo sample
to a linear combination of pure $\nu_e$~CC and $\nu_{e,\mu}$~NC
generated samples. The resulting $\nu_e$~CC-like fraction 
of $\nu_\tau$~CC events is 18\% for our data sample. 

The effect of $\nu_e \to \nu_\tau$ oscillations on the observed 
$\nu_e$ spectrum was determined in the following way:
a beamline simulation was used to tag the creation point of a
$\nu_e$ along the decay pipe giving the survival probablity for 
each $\nu_e$ as ($1- P(\nu_e \to \nu_\tau$)) from Eq.~(\ref{eq:posc}).
The predicted $\nu_e$ flux was normalized to the observed charged-current 
muon neutrino flux at the detector which was simulated in the 
same beamline Monte Carlo. We also added in the number
of $\nu_\tau$ charged-current interactions which 
would appear in the extracted $\nu_e$ sample at the detector.
The probablity of observing a $\nu_\tau$ charged-current interaction
in this data sample was calculated from the predicted normalized 
$\nu_e$ flux multiplied by the $\nu_\tau$ creation probability,
$P(\nu_e \to \nu_\tau$), and the $\nu_e$~CC-like fraction, $f_e$.
We took into account effects of $\nu_\tau$ charged-current 
cross section suppression by including mass suppression terms \cite{heavy},
kinematic suppression for massive particle production, and 
the altered visible energy spectrum ($E_{vis}$) for 
$\nu_\tau$ charged-current events which contains 
visible energy from the tau decay in determining the
effect of $\nu_\tau$ appearance.

\begin{figure}
\centerline{
\psfig{figure=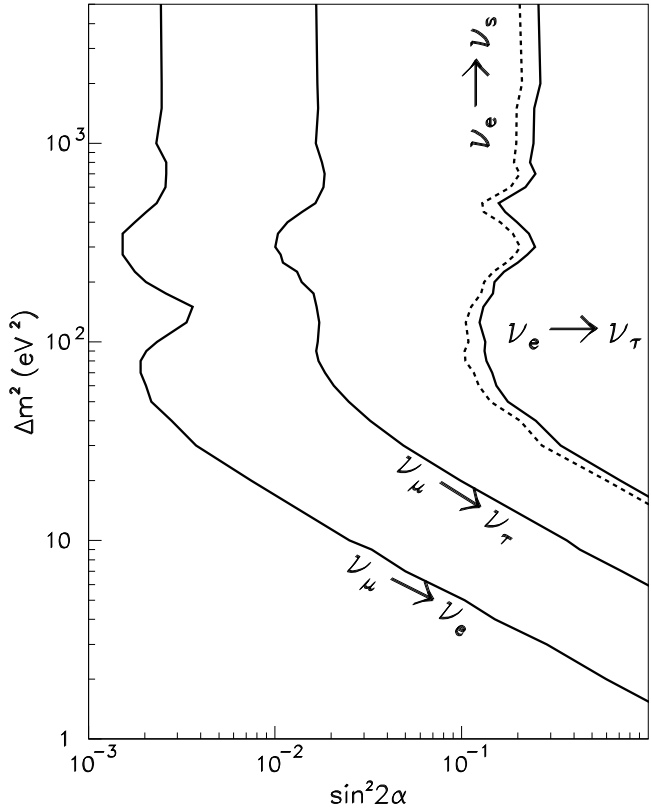,width=\columnwidth}}
\caption{Excluded region of $\sin^2 2\alpha$ and $\Delta m^2$
for (right to left) $\nu_e \rightarrow \nu_\tau$,
$\nu_e \rightarrow \nu_s$ (disappearance), 
$\nu_\mu \rightarrow \nu_\tau$,
and $\nu_\mu \rightarrow \nu_e$ (see Reference [7]) 
at 90\% confidence using the Feldman-Cousins approach.
The first three limits are new. The $\nu_\mu \rightarrow \nu_e$ 
limit differs from Reference [7]  
only in the construction of the 90\% confidence limit.}
\label{fig:fc}
\end{figure}

The effect of $\nu_\mu \to \nu_\tau$ oscillations on the $\nu_e$
spectrum depends on the creation probability of
$\nu_\tau$ from $\nu_\mu$, $P(\nu_\mu \to \nu_\tau$).
The $\nu_\tau$ appearance effect is calculated
by multiplying $P(\nu_\mu \to \nu_\tau$) by the
$\nu_e$~CC-like fraction, $f_e$ and weighting by 
the $\nu_\tau$~CC cross section suppression factor.
For completeness, we include a limit on $\nu_\mu \to \nu_\tau$
from this data sample (see Figure \ref{fig:fc}).

The major sources of uncertainties in the comparison of the $\nu_e$
flux extracted from the data to that predicted by the Monte Carlo are:
the statistical error from the fit in $\nu_e$ flux extraction,
error in shower shape modeling (described below),
uncertainty in the absolute energy calibration of the 
detector (1\%) which affects the relative neutrino flux extracted 
using a data sample with low hadron energy \cite{flux}, 
and finally the uncertainty in the predicted
flux of $\nu_e$'s at the detector which is estimated to 
be $4.1\%$ \cite{ca94}. This error is dominated by a 20\% production 
uncertainty in the $K_L$
content of the secondary beam which produces 16\% of the $\nu_e$ flux.
The majority of the $\nu_e$ flux comes from $K_{e_{3}}^\pm$ decays,
which are well-constrained by the observed $\nu_\mu$ spectrum from
$K_{\mu_{2}}^\pm$ decays \cite{ca94}.
Other sources of systematic errors were also
investigated and found to be small.

 The uncertainty in shower shape modeling is estimated 
by extracting the $\nu_e$ flux using two definitions of $\eta$.
Analogous to the definition of $\eta_3$ given in Eq.~(\ref{eq:eta3}),
we define $\eta_4$ to be the ratio of the sum of the energy deposited
outside the first {\em four} scintillation counters to the total visible
energy. If the modeling of the showers were correct, the difference in the
number of electron neutrinos measured by the two methods should be
small, any difference is used to estimate the systematic error. Since
this error was shown not to be correlated among energy bins, we add it
in quadrature to the statistical error from the fit.

\begin{table}
\caption{The result for $\sin^2 2\alpha$ from the fit at each $\Delta
m^2$ for $\nu_e \rightarrow \nu_\tau$ oscillations. The 90\% C.L.
upper limit is equal to the best fit $\sin^2 2\alpha +
1.28\sigma$.}
\label{tab:bestfit}
\begin{center}
\begin{tabular}{cccccc}
$\Delta m^2$ (eV$^2$) & Best fit & $\sigma$ & $\Delta m^2$ (eV$^2$) &
Best fit & $\sigma$ \\
\tableline
\begin{tabular}{r}
   10.0  \\
   20.0  \\
   30.0  \\
   40.0  \\
   50.0  \\
   60.0  \\
   70.0  \\
   80.0  \\
   90.0  \\
  100.0  \\
  125.0  \\
  150.0  \\
  175.0  \\
  200.0  \\
  225.0  \\
  250.0  \\
\end{tabular} & \begin{tabular}{r}
-0.775 \\
-0.208 \\
-0.112 \\
-0.046 \\
-0.045\\
-0.026\\
-0.012\\
-0.006\\
0.003\\
0.006\\
0.005\\
0.013\\
0.029\\
0.033\\
0.047\\
0.070\\
\end{tabular} & \begin{tabular}{r}
2.046 \\
0.553 \\
0.269 \\
0.180 \\
0.134 \\
0.109 \\
0.096 \\
0.086 \\
0.079 \\
0.078 \\
0.073 \\
0.072 \\
0.072 \\
0.071 \\
0.074 \\
0.080 \\
\end{tabular} & \begin{tabular}{r}
  275.0  \\
  300.0  \\
  350.0  \\
  400.0  \\
  450.0  \\
  500.0  \\
  600.0  \\
  700.0  \\
  800.0  \\
 1000.0  \\
 1500.0  \\
 2000.0  \\
 5000.0  \\
10000.0  \\
20000.0  \\
\end{tabular} & \begin{tabular}{r}
 0.083 \\
 0.094 \\
 0.065 \\
 0.025 \\
-0.011 \\
-0.024 \\
 0.045 \\
 0.065 \\
 0.045 \\
 0.040 \\
 0.045 \\
 0.062 \\
 0.053 \\
 0.060 \\
 0.048 \\
\end{tabular} & \begin{tabular}{r}
0.087 \\
0.089 \\
0.092 \\
0.098 \\
0.104 \\
0.111 \\
0.110 \\
0.112 \\
0.115 \\
0.124 \\
0.122 \\
0.124 \\
0.125 \\
0.124 \\
0.123 \\
\end{tabular}
\end{tabular}
\end{center}
\end{table}

The data are fit by forming a $\chi^2$ which incorporates the Monte
Carlo generated effect of oscillations and terms
with coefficients accounting for systematic uncertainties. A best fit
$\sin^2 2\alpha$ is determined for each $\Delta m^2$ by minimizing
the $\chi^2$ as a function of $\sin^2 2\alpha$ and these systematic
coefficients. At all $\Delta m^2$, the data are consistent with no
observed oscillations. Table \ref{tab:bestfit} shows the best fit value of
$\sin^2 2\alpha$ at each $\Delta m^2$ 
for $\nu_e \to \nu_\tau$ oscillations. The largest statistical
significance of a best-fit oscillation at any $\Delta m^2$ is 
$1\sigma$.

\begin{figure}
\centerline{
\psfig{figure=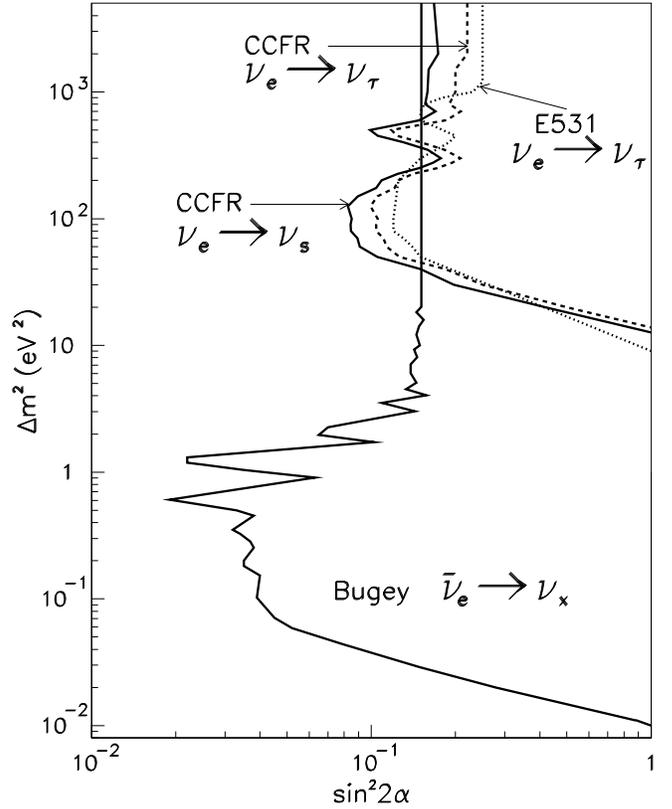,width=\columnwidth}}
\caption{Excluded region of $\sin^2 2\alpha$ and $\Delta m^2$ for
$\nu_e \rightarrow \nu_\tau$ and $\nu_e \rightarrow \nu_s$
oscillations from this analysis at 90\% confidence
(one-sided limit).}
\label{fig:osc}
\end{figure}

The frequentist approach \cite{pdg} is used to set a 90\% confidence
upper limit for each $\Delta m^2$. The limit in $\sin^2 2\alpha$
corresponds to a shift of 1.64 units in $\chi^2$ from the minimum
$\chi^2$ (at the best fit value in Table~\ref{tab:bestfit}). 
The 90\% confidence upper limit is plotted in Figure \ref{fig:osc} for
$\nu_e \rightarrow \nu_\tau$ and $\nu_e \rightarrow \nu_s$.
The best limits of $\sin^2 2\alpha$ are  
$< 9.9 \times 10^{-2}$ is at $\Delta m^2 = 125$~${\rm eV^2}$
and $< 8.3 \times 10^{-2}$ is at $\Delta m^2 = 125$~${\rm eV^2}$ 
respectively. For $\sin^2 2\alpha = 1$, $\Delta m^2 > 20$~${\rm eV^2}$ 
is excluded, and 
$\sin^2 2\alpha > 0.21$ for $\Delta m^2 \gg 1000$~${\rm eV^2}$
for $\nu_e \rightarrow \nu_\tau$.

As an alternative statistical treatment of this result
we present 90\% confidence limits based on the
unified approach of Feldman and Cousins \cite{fc}
recently adopted by the PDG \cite{pdg98}.
Figure \ref{fig:fc} shows all CCFR limits obtained 
using the longitudinal shower-shape method.
Our previously published limit \cite{alex} on
$\nu_\mu \rightarrow \nu_e$ 
used a one-sided confidence limit approach (as above). 

In conclusion, we have used a high-statistics sample of 
$\nu_e$ charged-current interactions in the CCFR
coarse-grained calorimetric detector to search for 
$\nu_e \rightarrow \nu_\tau$, $\nu_e \rightarrow \nu_s$
and $\nu_\mu \rightarrow \nu_\tau$ oscillations.
We see a result consistent with
no neutrino oscillations and find 90\% confidence level excluded
regions in $\sin^2 2\alpha - \Delta m^2$ phase space. 
This result improves on existing limits for $\nu_e \rightarrow \nu_\tau$
in the range $50$~eV$^2<\Delta m^2<200$~eV$^2$.

\end{document}